\documentclass[figures,cite]{epl}
\usepackage{graphicx}
\usepackage{amsmath}
\usepackage{psfrag}

\newcommand{\dd}{\text{d}}

\newcommand{\ee}{\text{e}}

\newcommand{\bv}{\text{\bf v}}

\usepackage[latin1]{inputenc}
\usepackage[T1]{fontenc}
\usepackage{ae,aecompl}

\begin{document}
\title{Injected power and entropy flow in a heated granular gas}
\author{P. Visco\inst{1,2} \and A. Puglisi\inst{1} \and A.
  Barrat\inst{1} \and E. Trizac\inst{2} \and F. van Wijland\inst{1,3}}
\shortauthor{P. Visco \etal} 
\institute{ 
  \inst{1}Laboratoire de
  Physique Th\'eorique (CNRS UMR8627) - B\^atiment 210,
  Universit\'e Paris-Sud, 91405 Orsay cedex, France\\
  \inst{2}Laboratoire de Physique Th\'eorique et Mod\`eles
  Statistiques (CNRS UMR8626) - B\^atiment 100,
  Universit\'e Paris-Sud, 91405 Orsay cedex, France\\
  \inst{3}P\^ole Mati\`ere et Syst\`emes Complexes (CNRS UMR7057) -
  Universit\'e Denis Diderot (Paris VII), 2 place Jussieu, 75251 Paris
  cedex 05, France
} 

\maketitle
\begin{abstract}
  Our interest goes to the power injected in a heated granular gas and to the
  possibility to interpret it in terms of entropy flow. We numerically
  determine the distribution of the injected power by means of Monte-Carlo
  simulations. Then, we provide a kinetic theory approach to the computation
  of such a distribution function. Finally, after showing why the
  injected power does not satisfy a Fluctuation Relation {\em \`a la}
    Gallavotti-Cohen, we put forward a new quantity which does fulfill such a
    relation, and is not only applicable in a variety of frameworks outside
    the granular world, but also experimentally accessible.
\end{abstract}

In the realm of non-equilibrium statistical physics, universal or
generic results are scarce. Landmarks in the field are Einstein's
seminal contribution~\cite{einstein} exhibiting a relation between
particle current fluctuations and a response function, followed by
Onsager's reciprocity \cite{onsager} and the Green--Kubo
relations~\cite{green,kubo}. The recent discovery by Evans, Cohen and
Morris~\cite{evanscohenmorris} formalized into a mathematical theorem
of general scope by Gallavotti and Cohen~\cite{gallavotticohen} of a
symmetry property, that we will call Gallavotti--Cohen relation (GC),
bearing on the entropy flux distribution therefore stands as a major
progress.
Similar results have been established for Markov
processes by Kurchan~\cite{kurchan} and Lebowitz and
Spohn~\cite{lebowitzspohn}, with much lighter mathematics than for
dynamical systems. However, mathematical as well as physical
difficulties, that we shall later describe, have prevented this result
to find its way towards experimental confirmation.

From the point of view of probing the validity of the
Gallavotti--Cohen theorem outside its mathematical domain of validity,
granular gases will prove instrumental. This was indeed already
realized by Auma\^{\i}tre {\it et al.}~\cite{aumaitrefauvemcnamarapoggi}.  
Yet, in
the absence of a first-principle definition of entropy, a physical
one, albeit heuristic, was proposed~: the energy injected by the
thermostat divided by the granular temperature.  This is precisely the
quantity simulated in~\cite{aumaitrefauvemcnamarapoggi}, and considered in a
recent experimental work~\cite{feitosamenon}. Besides, 
the idea of using a macroscopic and global
observable for characterizing the state of a system, instead of
resorting to local probes (velocity field, correlations, structure
factor) has proved a valuable tool {\it per se} for the
identification of generic features in non-equilibrium
systems~\cite{refpinton}.

In this letter we consider a paradigmatic model of a granular gas
maintained in a steady-state by external heating, namely hard-sphere
(or hard-disk) particles undergoing inelastic collisions, each of
these particles being independently subjected to a random force
${\bf F}_i$
which we take to be Gaussian distributed with variance $\Gamma$.
The equation of motion for particle $i$ with velocity 
$\bv_i$ then reads $d\bv_i/dt = {\bf F}_{\text{coll}}+ {\bf F}_i$ where
${\bf F}_{\text{coll}}$ is the force due to inter-particle collisions,
and acts at contact only \cite{referencesgazchauffe}.
This model
preserves essential features of standard experimental setups, like the
inelastic collisions, and a heating mechanism independent of the
particles' velocities. It moreover bypasses experimental difficulties
like the lack of translational invariance and has been the subject of
intense investigations both on the numerical and analytical
sides~\cite{referencesgazchauffe}, therefore its putative limitations
are controlled.

The results we have obtained are as follows: First, by means of Monte-Carlo
simulations, we pinpoint the numerical hazards that pave the way to the full
determination of the injected power probability distribution function (pdf)
and its symmetry properties. Second, we show how kinetic theory can be
extended to describe the large deviation function of the injected power thus
allowing for an explicit test of GC. This constitutes the first analytical
result of a distribution function for a global observable in a many-body, non
Gaussian, far from equilibrium system. Finally, we explain the reasons why the
injected power cannot satisfy a Gallavotti-Cohen relation. This leads us
to proposing a new quantity, with the properties of a Gibbs entropy
  flow, that not only avoids the aforementioned difficulties, but that
should also be accessible in specific granular gases experiments, and could
possibly be generalized to other non-equilibrium systems.

We define the time-integrated injected power over the time interval
$[0,t]$ as the total work ${\cal W}(t)$ provided by the thermostat:
\begin{equation} \label{work}
{\cal W}(t)=\int_0^t \dd t\;\sum_i {\bf F}_i\cdot \bv_i.
\end{equation}
The granular temperature is
defined by $T_g=1/\beta=\langle\bv_i^2\rangle/d$
where $d$ is the space dimension and the angular brackets denote a
statistical average in the steady-state. The central object of our
study is the pdf of ${\cal W}(t)$, $P({\cal W},t)$,
and its related large deviation function
\begin{equation}
\pi_\infty(w) =\lim_{t \to\infty} \pi_t(w) \quad\hbox{with }\quad \,\,
\pi_t(w) = 
\frac{1}{t}\ln P(w t,t).
\end{equation}
The 
generating function $\hat{P}(\lambda,t)=\langle\ee^{-\lambda {\cal
W}}\rangle $ and its related large deviation function
$\mu(\lambda)=\lim_{t \to\infty}\frac{1}{t} \ln \hat{P}(\lambda,t)$ will
provide equivalent information, as we may usually go from one to the other by
Legendre transform, $\pi_\infty(w)=\text{max}_\lambda\{\mu(\lambda)+\lambda
w\}$.

We now recall the content of the GC relation. In phenomenological
thermodynamics~\cite{degrootmazur}, one writes an evolution equation
for the entropy $S$,
\begin{equation}
\frac{\dd S}{\dd t}=\sigma-
\int_{\text{system}}\!\!\!\!\!\!\!\!\!\!\!\!\dd V\nabla\cdot{\bf J}_S
\end{equation}
which, in a steady-state, expresses a balance of an entropy
production term (a measure of intrinsic irreversibility) with an
entropy flux ${\bf J}_S$ produced by an external drive (typically, a
boundary condition imposing energy or matter flow). The central result
of Gallavotti and Cohen is to have shown in a particular setting that
the pdf $P({\cal S},t)$ of a quantity ${\cal S}(t)$ playing the role of the
time-integrated flux term $\int\dd t \dd V\nabla\cdot{\bf J}_S$ verifies, in the limit of
asymptotically large times,
\begin{equation}\label{GC1}
\ln \frac{P({\cal S},t)}{P(-{\cal S},t)}={\cal S}.
\end{equation}
Nevertheless, the assumptions underlying GC are microscopic
reversibility and the identification of $\dot{\cal S}$ with the phase
space contraction rate. In granular materials, the former is violated
and the latter definition is doomed to fail, since the phase space
volume can only decrease if the heating mechanism is velocity
independent. This has led to the idea that $\beta
{\cal W}$ could possibly play the rôle of ${\cal S}$ 
\cite{aumaitrefauvemcnamarapoggi}. In the heated granular
gas the fluctuating total kinetic energy
$E(t)=\sum_i{\bv_i}^2/2$ varies according to
\begin{equation}\label{fluctuatingbalance}
E(t)-E(0)={\cal W}(t)-{\cal D}(t)
\end{equation}
where ${\cal D}(t)\geq 0$ is the energy dissipated through collisions
up until time $t$. Of course on average $\langle{\cal
W}(t)\rangle=\langle{\cal D}(t)\rangle=2d\Gamma t>0$, yet there will be
individual realizations for which ${\cal W}$ will be negative. If the
GC relation (\ref{GC1}) held for ${\cal W}$, it would take the
following form \cite{rque2}
\begin{equation}\label{GC2}
\pi_\infty(w)-\pi_\infty(-w)=\beta w,\;\mu(\lambda)=\mu(\beta-\lambda)
\end{equation}
Remark, though, that the left-hand side in
(\ref{fluctuatingbalance}) is bounded in time (while
${\cal W}$ and ${\cal D}$ grow typically linearly with time), 
therefore~\cite{farago2} both
the injected power and the dissipated power have the same (cumulant)
generating 
function. A straightforward consequence of that fact is the absence of $w<0$
events at large times: \begin{math}\pi_\infty(w<0)=-\infty\end{math}, hence (\ref{GC2}) cannot be
correct \cite{rque}. 

Let us see now how this rigorous fact comes about in numerical studies. We
have simulated a model of $N$ inelastic hard disks under the effect of a
homogeneous thermostat. The collisions between two disks $i$ and $j$ conserve
the total momentum and reduce the normal component of the relative velocity,
i.e.  $(\mathbf{v}_i'-\mathbf{v}_j')\cdot\mathbf{\hat{n}}=-\alpha
(\mathbf{v}_i-\mathbf{v}_j)\cdot\mathbf{\hat{n}}$ where the primes mark the
post-collisional velocities and $\mathbf{\hat{n}}$ is the direction joining
the colliding particles.  The driving force ${\bf F}_i$ mentioned above obeys
$\langle F_i^\gamma(t)F_j^\delta(t')\rangle=
2\Gamma\delta_{ij}\delta^{\gamma\delta}\delta(t-t')$.  Molecular Chaos is
enforced through the use of a Direct Simulation Monte Carlo~\cite{dsmc}
algorithm. The restitution coefficient $\alpha$ takes values between $0$ and
$1$ (elastic gas). The gas rapidly reaches a stationary state with a granular
temperature $T_g=4\Gamma/(1-\alpha^2)$, where the mean free time is used as a
time unit, which will be the case in the subsequent analysis. We have measured
the fluctuations of the work exerted by the thermostat ${\cal W}(t)$ as
defined in Eq.~(\ref{work}) using an integration time $t$.  $P({\cal W},t)$
clearly shows non-Gaussian tails, which can be regarded as a hint that the
true large deviations are being probed (since central limit theorem requires
small deviations to be Gaussian).  The resulting function $\pi_t(w)$ is shown
(see Fig.~\ref{fig:pdf}) to verify almost perfectly a relation such as
(\ref{GC2}). When decreasing $\alpha$ a slight departure (about $10\%$) from
the slope $\beta$ can be recognized. And yet the plot of $\pi_t(w)-\pi_t(-w)$
stays linear in $w$.  The inset of Fig.~\ref{fig:pdf} is of particular
interest: it is numerically the proof that the slope does not depend upon the
time of integration for the considered range of $t$ values.  More
interestingly, it highlights the dramatic decay of the number of observable
negative events as $t$ is increased: even with such a simple and fast to
simulate model, obtaining large statistics at large $t$ is difficult.  This
observation poses a crucial question: has the numerical investigation reached
the true asymptotics, i.e.  can we believe that $\pi_t(w) \sim \pi_\infty(w)$
for the values of $t$ shown in the figure? In the following we exploit further
numerical studies and a novel analytical approach to demonstrate that a time
much longer should be waited and that this first numerical result is
misleading.  It is nevertheless important to discover that at values of $t$
small enough to observe some negative injected power the GC relation appears
to be satisfied.

We now sketch the analytical strategy that allowed us to find the pdf of
the injected power.  We start from an extended Liouville equation for the
probability $\rho(\Gamma_N,{\cal W},t)$ that the system is in state $\Gamma_N$
with ${\cal W}(t)={\cal W}$ at time $t$. The second step is to convert the
Liouville equation in terms of the generating function
$\hat{\rho}(\Gamma_N,\lambda,t)=\int\dd{\cal W}\ee^{-\lambda{\cal W}}\rho$.
Recall that, for $\lambda=0$, projection of the Liouville equation onto the
one-particle subspace yields an equation coupling the one-particle
distribution function to the two-particle distribution function, which is
factorized through the molecular-chaos hypothesis. In our case, the situation
is very similar when it comes to determining the largest eigenvalue
$\mu(\lambda)$ of the evolution operator for $\hat{\rho}$ (which, in physical
words, is the generating function for the cumulants of ${\cal W}(t)/t$):
projecting onto the one-particle subspace yields an equation coupled to the
two-particle subspace.  By means of a molecular-chaos like closure procedure
we arrive at an equation for both the eigenvalue $\mu$ and its related
eigenfunction. The extensive mathematical analysis of this program 
is cumbersome and will be
reported elsewhere~\cite{viscopuglisibarrattrizacvanwijland}. 
We concentrate here on the results that are
as follows. First we find that the large deviation function of ${\cal W}= w t$
has the graph depicted in Fig.~\ref{fig:mu}.  The tails are given by
\begin{equation}
\pi_\infty(w\to 0^+)\sim -w^{-1/3},\;\;\pi_\infty(w\to+\infty)\sim -w
\end{equation}
with, as expected, no $w<0$ contribution. While the $w\to 0^+$ regime appears
to be thermostat-dependent, the exponential right tail of $P({\cal W},t)$
seems to be a robust property related to the presence of a branch cut in the
complex $\lambda$ plane for $\mu(\lambda)$. 

\begin{figure}[t]
\twofigures[width=8cm]{gcnew.eps}{pidiq_2.eps}
\caption{{\bf Color online}: 
  Plot of $\pi_t(w)-\pi_t(-w)$ as a function of $\beta w$.  The dashed line
  has slope $1$, the dotted line has slope $1.1$. The inset contains the same
  graph for different values of $t$, for the case $\alpha=0.9$, $N=100$ and
  $\Gamma=0.5$.}
\label{fig:pdf} 
\caption{The large deviation function $\pi_\infty$ of the rescaled 
  quantity ${\cal W}/ \langle {\cal W} \rangle$ obtained from the Legendre
  transform of the largest eigenvalue $\mu (\lambda)$ (see text and
  \cite{viscopuglisibarrattrizacvanwijland} for details). The inset shows the
  dimensionless third cumulant $\langle{\tilde {\cal W}}^3 \rangle_c = \langle
  {\cal W}^3 \rangle_c \beta^2 / \langle {\cal W} \rangle$ relaxing towards
  its predicted theoretical value, which is~$8$.  The part of the curve close
  to $w/\langle w \rangle$ is magnified to show the agreement between
  simulations and theory on the limited accessible range of values at $t=40$.}
\label{fig:mu}
\end{figure}

We now come back to our previous numerical results. The observation of a clean
linear behaviour for $\pi_t(w)-\pi_t(-w)$, with slope $\beta$, apparently
independently of time (see inset of Fig.~\ref{fig:pdf}), as though GC were
satisfied, is in contrast with our analytical results indicating that the
truly asymptotic $\pi_\infty(w)$ has no $w<0$ contribution.  At the level of
simulations or experiments it is impossible to distinguish between
non-Gaussian tails due to spurious short time effects and real asymptotic
large deviation tails.  Nevertheless the numerical study of cumulants with
time, which are integrals and therefore benefit from a larger statistics, is
decisive both to prove the validity of our analytical approach and to get rid
of all doubts about which is the true asymptotic regime.  We see (inset of
fig.~\ref{fig:mu}) that the third cumulant reaches a stable value at times of
order $\sim 50$, much larger than the times used in Fig.~\ref{fig:pdf}.
Moreover this stable value is in very good agreement with the value expected
from our theory $\langle{\cal W}^3 \rangle_c
=-t\,[\partial^3\mu(\lambda)/\partial \lambda^3]_{\lambda=0}$. We recall that
this estimate of the third cumulant is highly non-trivial. At such large $t$
values, however, the numerically accessible range of $w$ is very small,
limiting the possibility of a detailed comparison between analytics and
numerics (see Fig.  \ref{fig:mu}).

Having reached the conclusion that the injected power cannot fulfill the GC
relation, in this final paragraph we propose an alternative quantity which is,
by its very definition, a Gibbs entropy flow, and satisfies the GC fluctuation
relation.  The sequel applies both to the original experimental system in
which particles contained in a closed box are vigorously shaken and to our
random thermostat.  We now tag one of these particles that we follow along its
path, which is similar in spirit to the work reported in
\cite{martinpiasecki}. The rest of the particles act as a thermostat for the
tagged particle. Note however that the states of the particle bath evolve
according to dynamical rules that do not fulfill the detailed balance
condition. Similarly, the trajectory in the phase space of the tagged particle
does not fulfill the detailed balance condition (this is a key difference with
\cite{martinpiasecki}).  It is now possible to view the time evolution of the
tagged particle as a Markov process. In the tagged particle phase space, to
every non-zero transition rate between two velocity states one can associate a
non-zero rate for the backward move.  This is the core of the difference with
following the dynamics in the phase space of the whole system. There, due to
irreversible microscopic dynamics (inelastic collisions), for a given
forward-in-time trajectory in phase space, there is no corresponding, however
extremely unlikely, backward-in-time trajectory.  Hence for the tagged
particle it is perfectly legitimate to define an integrated entropy flow {\it
  \`a la} Lebowitz and Spohn~\cite{lebowitzspohn} (see also Maes~\cite{maes} or
  Gaspard~\cite{gaspard} for alternative presentations):
\begin{equation}\label{entropyflow}
{\cal S}(t)=\sum_{i=1}^{n(t)}\,\,
\ln\,\frac{K(\bv_i\to\bv_{i+1})}{K(\bv_{i+1}\to\bv_{i})}
\end{equation}
where $n(t)$ is the number of collisions undergone by the tagged particle over
the time interval $[0,t]$, $\bv_i$ is its velocity after the $i$-th collision,
and $K(\bv\to\bv')$ is the velocity transition rate of a particle undergoing a
collision. Expression (\ref{entropyflow}) may be experimentally accessible in
the following way: the monitoring of a large number of collisions (and of the
velocities before and after collision) first gives an estimation of the
velocity transition rates $K(\bv\to\bv')$, that can be tabulated with an
appropriate discretization of velocities; it is then possible to follow one
given tagged particle, to measure its successive pre- and post-collision
velocities and thus to compute ${\cal S}$. In the elastic limit for the
thermostat the transition rates $K(\bv\to\bv')$ verify the detailed balance
condition with respect to a Maxwellian tagged particle velocity
pdf~\cite{martinpiasecki}, hence there is no entropy flow anymore, and ${\cal
  S}(t)/t\stackrel{t\to\infty}\to 0$ (the tagged particle is in equilibrium
within its own phase space). Fully explicit expressions for the transition
rate $K(\bv\to\bv')$, and hence for ${\cal
  S}(t)$~\cite{viscopuglisibarrattrizacvanwijland} can be derived for an
inelastic thermostat, {\it e.g.} using Sonine expansions. By construction, the
pdf of the instantaneous entropy flow (\ref{entropyflow}) will verify a GC
relation, with the ensuing consequences in terms of Green--Kubo relations.

We have provided the first computation of the large deviation function of the
injected power in a system driven far from equilibrium, for which no general
theory has hitherto been coined, as opposed to systems weakly driven out of
equilibrium~\cite{derridabodineau}. This computation allowed us to pinpoint
numerical limitations and yet to extract reliable data. We have shown why the
injected power cannot satisfy a GC relation, thus revealing a scenario
sharing many features in common with the equilibrium toy model considered by
Farago~\cite{farago1}.  However, even if $\pi_\infty$, the large $t$ limit of
$\pi_t(w)$, cannot exhibit a GC symmetry due to the absence of a negative
tail, it appears that from a practical point of view, $\pi_t(w)$ seems to obey
a GC-like relation (see Fig \ref{fig:pdf}).  An important consequence of the
analytical work outlined here is that such a ``finite-time'' property cannot
be considered as an extension of GC theorem.  Finally we have put forward a
new approach, based upon a Lagrangian point of view, that leads to a
definition of an entropy flow possessing all the properties requested by the
systematic approach of Lebowitz and Spohn~\cite{lebowitzspohn}, and which may
be experimentally accessible. We are confident that not only the new
extensions of kinetic theory developed here, but also the proposal for an
entropy function, will trigger new theoretical investigations and will lend
themselves to experimental confirmations far beyond the granular community, in
the spirit of recent experiments on turbulent flows resting on Lagrangian
viewpoint investigations~\cite{lagrangianviewpoint}.


\acknowledgements 
The authors acknowledge several useful conversations with S.
Auma\^{\i}tre, J. Farago and S. Fauve.  A. P. acknowledges the Marie
Curie grant No.  MEIF-CT-2003-500944.  E.T. thanks the EC Human
Potential program under contract HPRN-CT-2002-00307 (DYGLAGEMEN).


\end{document}